\documentclass{article} 
\usepackage{graphicx}
\begin{document}

\centerline{\Large \bf 100 million years after the Big Bang}

\centerline{\bf Jeremy Mould}
\centerline{\bf Centre for Astrophysics \& Supercomputing, Swinburne University}

\begin{abstract}
Dark Energy Camera on the Blanco 4 meter telescope not only has the focal plane size the 4 meters were built for,
but also has excellent near infrared response. A DECam Deep Fields program is outlined,
which can reach M* galaxies at redshift 6 at a wavelength of one micron.
What reionized the Universe, when did globular clusters form, were there very massive stars and how did they end, and how did supermassive black holes emerge
a few hundred million years after the Big Bang ? These are some of the questions wide field high z surveys
in the infrared will open to observational study.
\end{abstract}

\section{Introduction to the EoR}
Until around 400 million years after the Big Bang, the Universe was a very
dark place. There were no stars, and there were no galaxies.  But is that
early epoch relevant to this meeting? Clearly yes, because the meeting is
about wide field astronomy. Also, it's about bulges, and so relevant unless
you think our bulge is 100\% pseudobulge and started forming after a billion
years, rather than 10$^8$ years. This question is an open one with some
persuasive dynamical evidence and some persuasive stellar populations
evidence, which don't seem to agree on the nature of the bulge. Bulge
classification has bifurcated (Kormendy \& Kennicutt 2004). There are now
classical bulges (formed at high redshift via mergers/accretion/rapid collapse
of overdensity) and pseudobulges (formed by secular erosion of disks). The
Local Group context is in Table 1. A review of the M31 bulge is given by Mould
(2013).

\vspace{0.5 cm}
\leftline{\bf Table 1: Local group bulges}
\begin{tabbing}
Galaxy sssss\= Bulge/disks \= Refssssssss       \= MBHsssss      \= Ref\kill
Galaxy \>Bulge/disk  \>Ref        \>M$_{BH}$(M$_\odot$)   \>   Ref\\
M31    \>0.35        \>WPS03      \>1.4 x 10$^8$\>   B+05\\
Milky Way\> 0.15     \>BS80       \>3 x 10$^6$  \>     G+05\\
M33    \>0.03        \>RV94       \>$<$1500     \> G+01\\
M32    \>$>$1          \>G01        \>3.4 x 10$^6$\>   vdM+98\\
\end{tabbing}

This contribution focuses on classical bulges and their possible formation in the first few hundred million years. A 10 M$_\odot$ star forming 100 Myr after the Big Bang would be seen today with a Balmer Jump at 11$\mu$m and a Lyman limit at 2.7$\mu$m. Such early stars, if they formed at all, are JWST targets.

One of the scariest problems in probing the high redshift Universe is the rapid rise of luminosity distance with redshift. In luminosity distance the epoch of reionization (EoR) from t = 100 Myrs extends 80 Hubble radii. We have spent the best part of the 20th century getting a passable picture of one Hubble radius. More powerful telescopes and instruments are needed for the EoR. Here is the list rating a mention here:
\begin{itemize}
\item JWST: powerful, but small field, with spectroscopy paramount
\item DECam: 1$\mu$ and shorter, wide field
\item KDUST: 1$\mu~<~\lambda~< 3\mu$, wide field, IR camera TBD, optical GPC AO published
\item Las Campanas Transit Survey: wide field ($\S 4$)
\item TMT high resolution (not elaborated here)
\item SKA: redshifted neutral hydrogen
\end{itemize}
Since this is CTIO's 50th anniversary, DECam will be the focus of this contribution.

\section{DECam Deep Fields}
Our team has commenced work on three fields with the goal: what is doing the reionizing in the EoR ? A galaxy luminosity function for deep fields is shown in Figure 1.
All 3 are circumpolar.

\begin{itemize}
\item Chandra Deep Field South (little data)
\item Prime Field (1.5 nights' data)
\item 16h field (no data)
\end{itemize}
How deep are we going ? For reference, M* at z = 6 is Y = 24.0 mag.
With the DECam deep fields we are not trying to compete with Hubble
(e.g. Illingworth et al 2013). In their XDF
all WFC3 near-IR and optical ACS data sets have been fully combined and accurately matched, resulting in the deepest imaging ever taken at these wavelengths ranging from 29.1 to 30.3 AB mag (5$\sigma$ in a 0.35$^{\prime\prime}$ diameter aperture) in 9 filters. The DECam deep fields are competitive in volume at z = 6, $not$ in attaining the highest redshift. An extract from our Prime Field is shown in Figure~2.

\begin{figure}[h]
\begin{center}
\includegraphics[angle=-0,width=.8\textwidth]{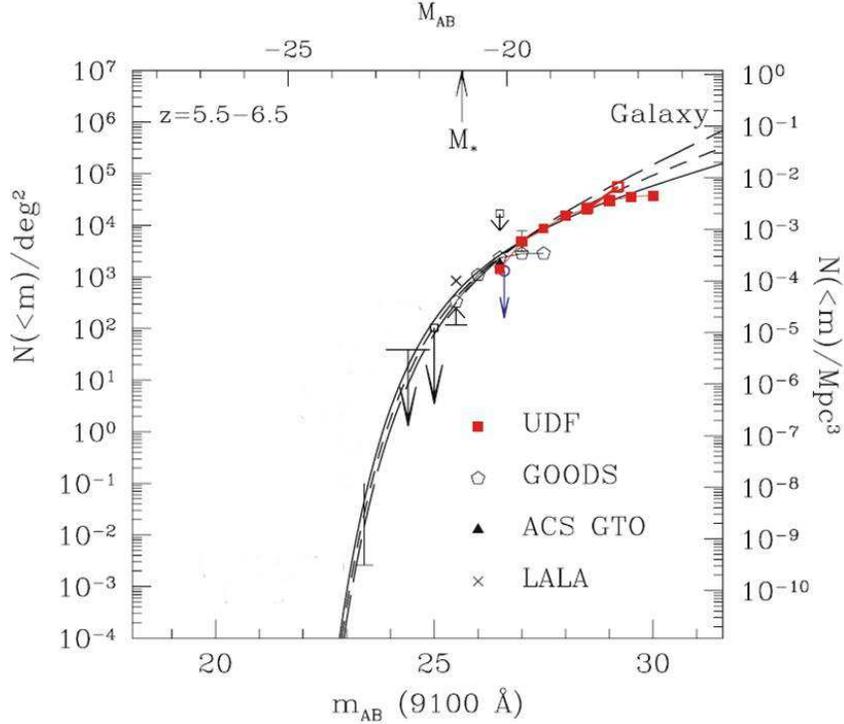}
\end{center}
\vspace{-.35 truein}
\caption{Galaxy counts at 910 nm according to existing deep fields for z $\sim$ 6.
AB mags at 910 nm $\approx~ z$ (Beckwith et al 2006).}
\end{figure} 
\begin{figure}[h]
\vspace{-.1 truein}
\begin{center}
\includegraphics[angle=-0,width=.9\textwidth]{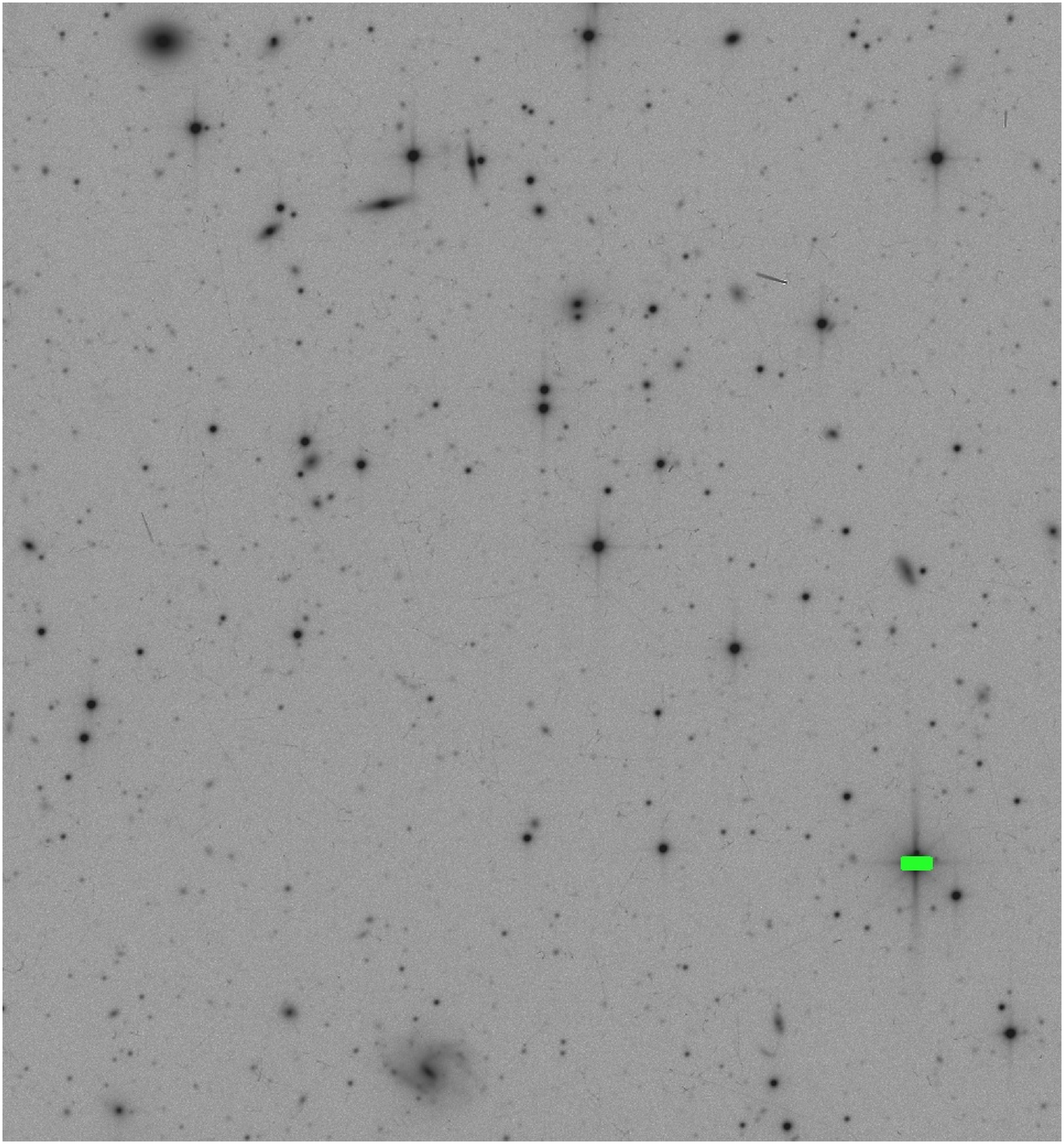}
\end{center}
\vspace{-.2 truein}
\caption{A few arcmin of the 140 minute stacked Y band image of the DECam Deep Fields Prime Field direct from the NOAO pipeline.}
\end{figure}

The DAOPHOT program (Stetson 1987) provides a CMD from aperture photometry (Figure 3). This photometry is calibrated with E region exposures. A provisional calibration for $z$ and Y is provided by selecting solar type standards from the E region fields and assuming that their I-$z$ and $z$-Y colors are those of a solar temperature blackbody at their effective wavelengths, with zero color corresponding to a Vega temperature blackbody. We expect to have reliable extinction coefficients after a year's experience of DECam observing. I dropouts are z = 6 candidates, and some examples are shown in Figure 4.

\begin{figure}[h]
\vspace{-.1 truein}
\begin{center}
\includegraphics[angle=-90,width=1.0\textwidth]{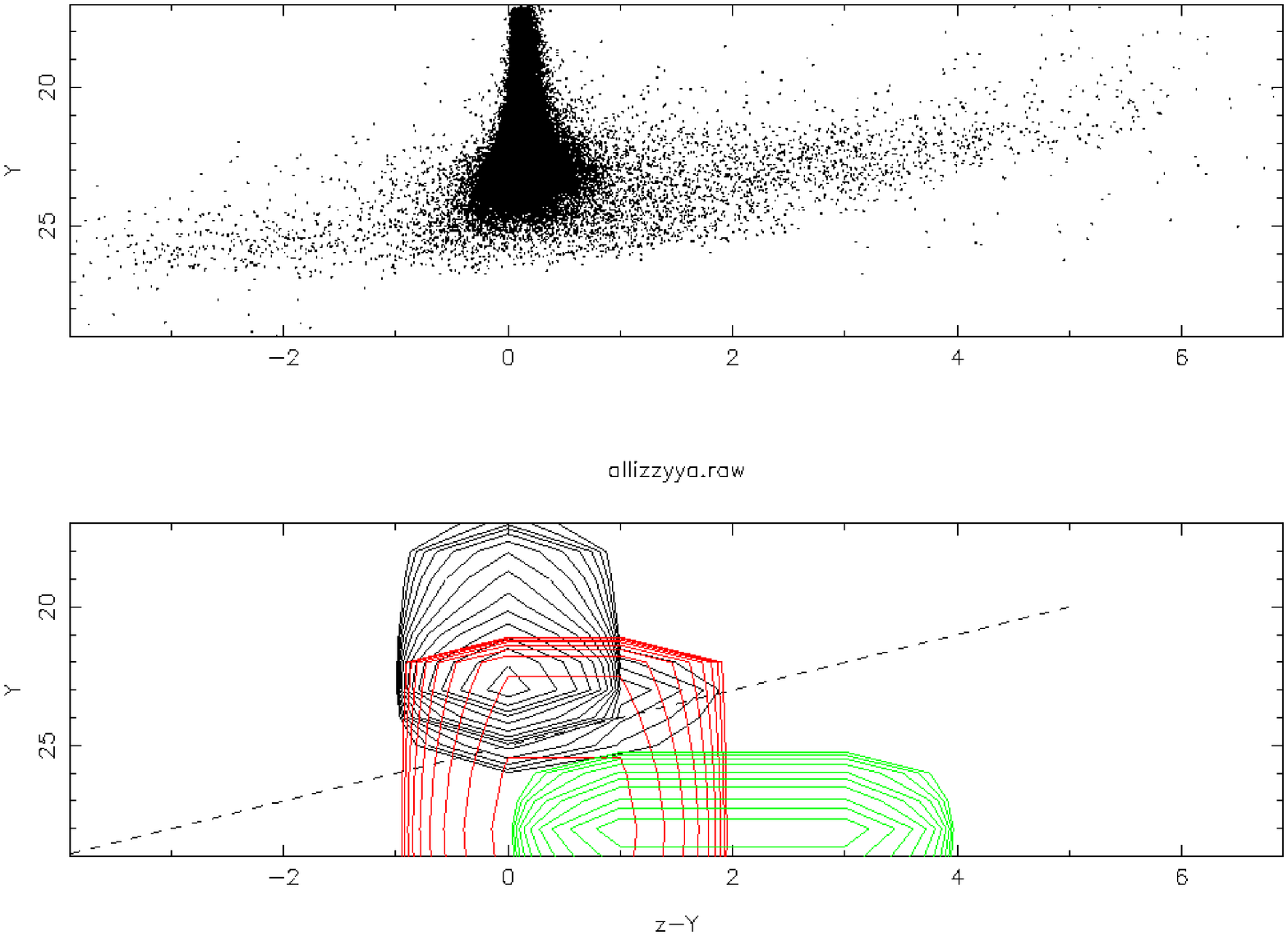}
\end{center}
\vspace{-.2 truein}
\caption{\small{CMD in the Prime field from our two 2012B nights. The red contours are star counts from the Bahcall-Soneira model of the Galaxy. The spacing is factors of $\surd$2. The green contours are galaxy counts at z = 5.5--6.5 from the Hubble UDF. The black contours are our counts. 
The dashed line is a completeness line at $z$ = 25 mag. We need to go a little deeper to better overlap the high redshift galaxy contours.}}
\end{figure} 


\begin{figure}[h]
\vspace{-.1 truein}
\begin{center}
\includegraphics[angle=-90,width=.99\textwidth]{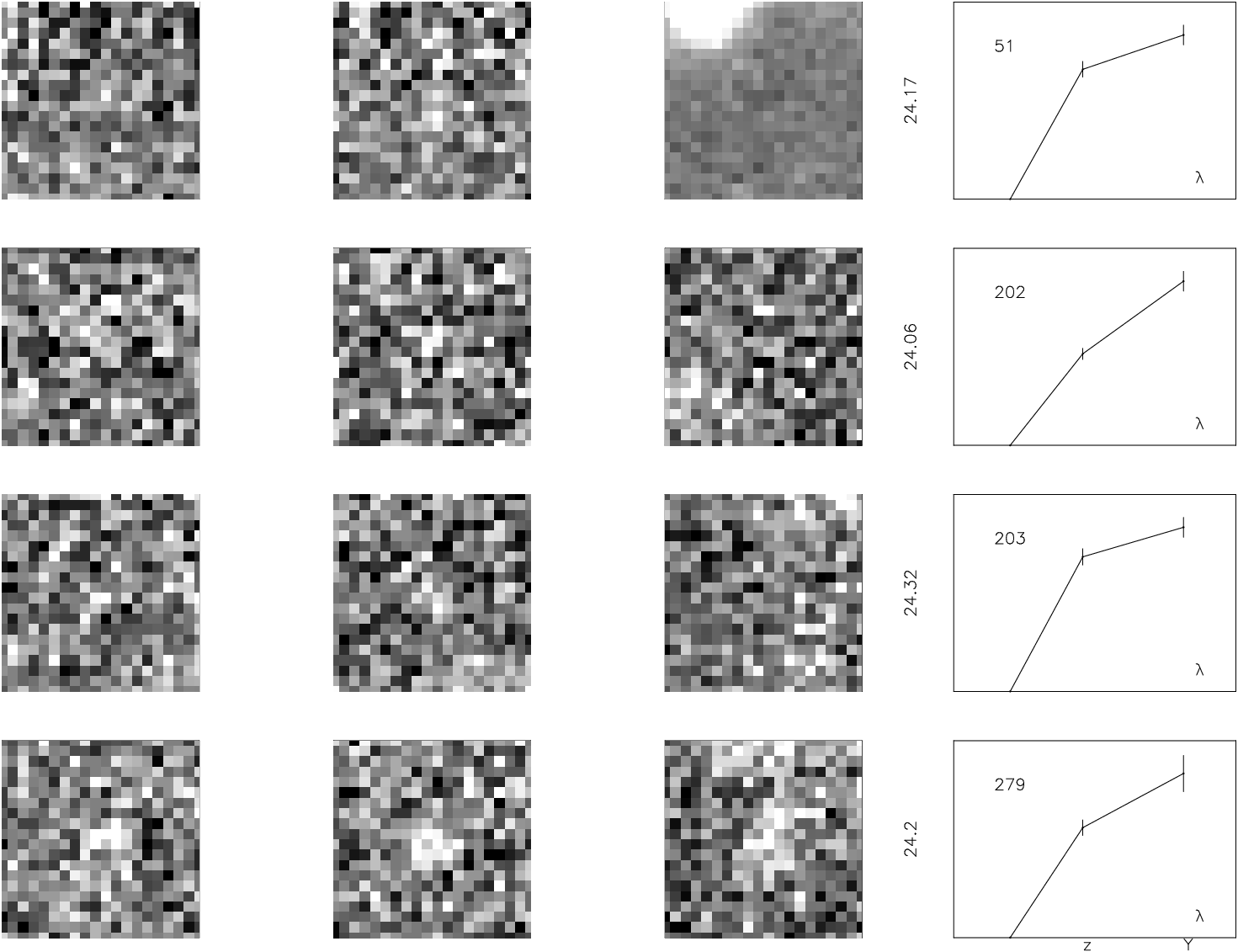}
\includegraphics[angle=-90,width=.99\textwidth]{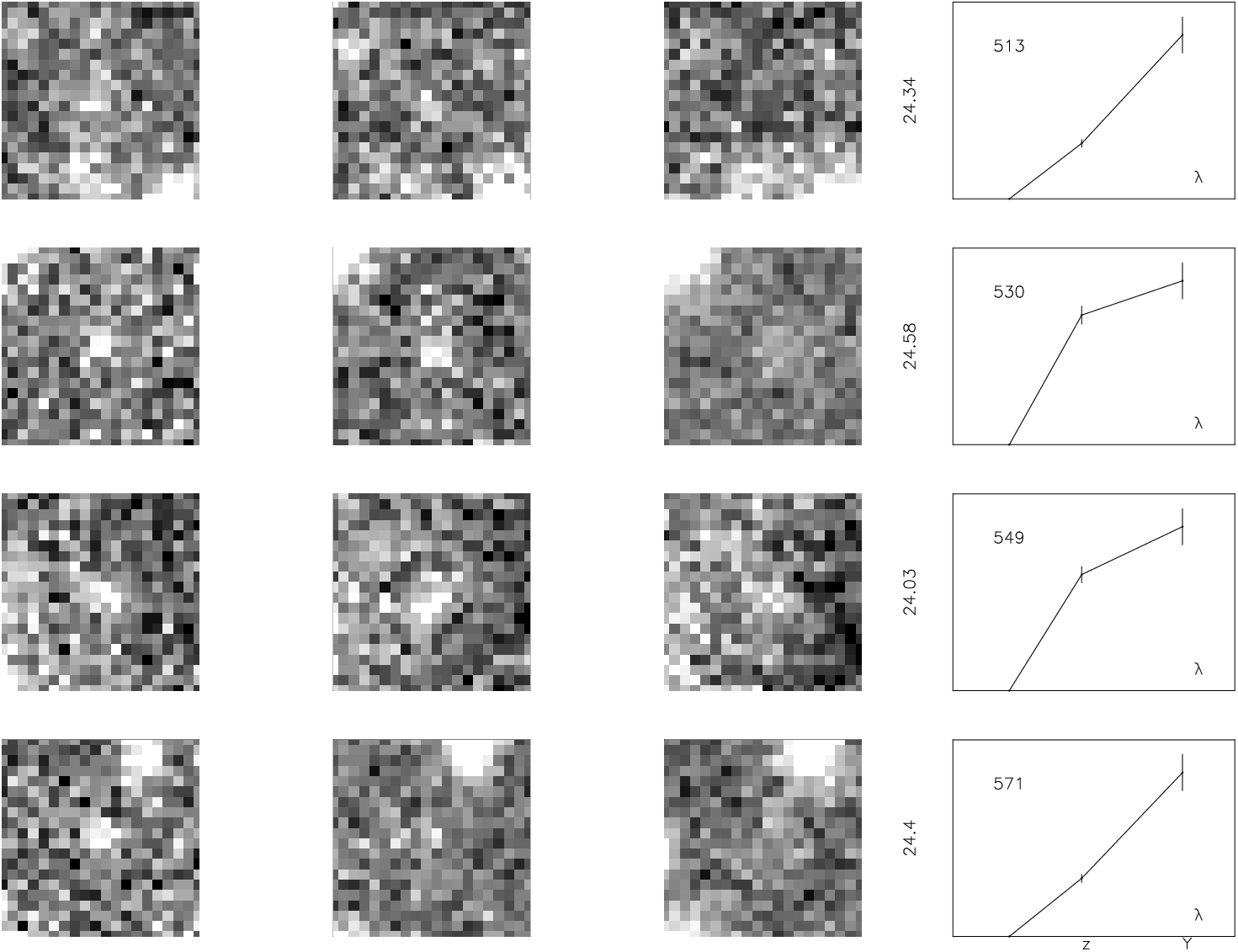}
\end{center}
\vspace{-.2 truein}
\caption{\small{Postage stamps of $i$-dropout candidates at 24th Y mag. The ordering l-to-r is Y,$z$,i, and a flux plot is shown at the right for each object. The Y mag is also shown.}}
\end{figure}

\subsection{Next steps DECam deep fields}
We have sought further time to complete these fields and bring them to 26th magnitude.
\begin{itemize}
\item With about twice the signal to noise we have here we'll have viable target lists for Gemini Flamingos redshifts
\item Relevant emission lines are  CIV etc all the way to [OII] $\lambda$3727
\item We also seek more time to monitor the deep fields for supernovae (Whalen et al 2013)
\item use the highly efficient Lyman break galaxy monitoring technique (Cooke 2012) to find z  $\sim$ 2--4 SLSNe) 
\item SLSNe will rise to peak from 10 -- 30 days, stay there for 2 -- 20 days, then decline in 20 -- 100 days. In the observer frame, this is 75 -- 230 day rise, 15 -- 150 days near peak, and 150 -- 750 day decline for objects at the mean redshift of z $\sim$ 6.5.
\item Expected rates: several per field
\item Pair Instability Supernovae might be an indicator of a top heavy initial mass function at high z.
\end{itemize}
There may also be a Supermassive SN chemical signature. SMS SNe produce very little nickel-56 (Heger et al. 2012) 
Their chemical signature may be distinguishable from those of most very massive (i.e. 140--260 M$_\odot$ ) Pop III explosions (PSNe), which can produce iron-group 
(Heger \& Woosley 2002) elements. Similar to these PSNe, however, SMS SNe would not make any s-process or r-process contributions.

\section{KDUST, formerly PILOT}
KDUST is a proposed Chinese Academy of Sciences 2.5m telescope at Dome A (http://kdust.org and Mould 2011). This facility would have significant advantages over existing 2$\mu$ survey telescopes, such as VISTA, whose survey work has been so impressive at this meeting. A science case for a very deep survey is under development and itemized in Table 2. Regarding PILOT, see Lawrence et al (2009).
The Antarctic advantages are
\begin{itemize}
\item almost diffraction limited images
\item wide field isoplanatism
\item low 2$\mu$m background
\end{itemize}
This combination is only available from the Antarctic plateau,
high altitude balloons and space.

\pagebreak
\leftline{\bf Table 2:  Prioritized science case for KDUST}						

\begin{tabbing}			
sssssssssssssssssssssssssssssssssssssssss\=Pri\=ssssField\=Competi\=Limt\=sssssMethod and comments\=Volu\=Dists\=Speeds\=Theme\kill
~~~~~~~~~~~~~~~~~~~~~~~~~~~~~~~~~~~~~~~~~~Priority\>\>	Field\>Comp-\>
K\>	Method and comments\>	Vol\>	Dist\>	Speed \>	Theme\\
\>\>size\>-etitor\>limit\>	\>Gpc$^3$\>\>ratio\>\\
Spatial variation galaxy LF at z = 6\>	B\>	$>>$HDF\>	VLT\>	25\>
HAWK-I has 7.5$^\prime$ field\>	\>\>		1\>	E\\
Weak lensing cosmology parameters\>	A\>	15000\>	Euclid\>
\>GPC\>			\>\>\>	D\\
IMF from 0.1 to 0.01 M$_\odot$\>	B\>	10 fields\>	VLT\>\>
see WFIRST science case\>	\>\>		1\>	G\\
Pair Instability SN at z $>$ 4\>	B\>	100\>	VLT\>	26\>
\>3.7\>	\>	1\>	T\\
Kuiper belt census and properties\>	C\>	20000\>	LSST/PS\>
\>GPC\>	\>\>\>			T\\
Cool white dwarfs \& the Milky Way\>	B\>	20 fields\>	VLT\>	27.5\>
see WFIRST science case\>\>		10* \>	1\>	G\\
Planetary transits\>	A\>\>		Kepler\>\>low scintillation
photometry\>			\>\>\>	T\\ 
Clusters of galaxies at z $>$ 2\>	A\>	100\>	SPT\>	26\>
followup redshifts helpful\>	\>\>2.3 \>			E\\
Lyman alpha emitters at z $>$ 9\>	?\>\>	VLT\>\>	 	narrowband\>
\>\>\>	E\\
Formation of the first SMBH\>	C\>	\>JWST\>\>		AGN at z $>$
6\>\>\>\>	E\\
Formation globular clusters at z $>$ 6\>	C\>\>	JWST\>\>
resolve galaxies at z $>$ 6\>\>\>\>	E\\
Y band dropouts at z = 10\>	B\>	100\>	VLT\>	26\>	combine w PSNe
survey\>	3.9 \>\>		1\>	E\\
\end{tabbing}									
									
\noindent Speed ratio is D/$\surd$(B), assuming no fov difference, where D is telescope
diameter and B is background.			\\				
GPC = Gigapixel CCD camera			\\				
``Theme" is evolving universe, dark universe, transient universe, galactic, see http://www.caastro.org\\
Anything VLT accessible is priority B, but that should be reassessed if 100
sq deg is really required for 3 of the projects\\	
Note that none of the present KDUST collaborators have ESO member VLT access\\	
Field is in sq deg except where stated otherwise.\\		
We assume the KDUST IR camera has a 8.5 arcmin field.\\				
Volume refers to a one mag range in luminosity distance\\			
The SDSS SN rate is 27000 SNe/yr/Gpc$^3$ (Dilday et al 2010). Massive star SNe
may be rarer than that by (m/10)$^{-4}$	simply from IMF considerations.\\	
*Units in the distance column are kpc.\\
WFIRST science case: Green et al (2011) and Spergel et al (2013).


\section{Las Campanas Transit Survey}

Off axis mirrors can be turned into transit telescopes for deep surveys to 25th mag. They can essentially be mounted in a static mirror cell and pointed at the zenith. For an f/2 GMT mirror a prime focus camera can be positioned on a 16 meter tower beside the mirror, pointing at the mirror center.
The first GMT mirror is already available for this purpose. If the AAO built a static mirror cell, and Carnegie transported the mirror to Las Campanas Observatory, it could be set up beside the Magellan Telescope. 
 Only the optics would need rudimentary protection from bad weather.
Storage charges in Tucson would be avoided.
The LCTS project could run the best part of a decade, repay early investment in the mirrors, build on the ANU's SkyMapper software system and yield many petabytes of unique data.
The proposal is not currently supported by the GMT Board. A design for off-axis corrective optics to flatten the focal plane has not yet  been attempted.

\section{Concluding thoughts}
One of the most powerful facilities for EoR science will be
the Square Kilometre Array (Taylor 2013). The long wavelength array to be built in Western Australia will show the first bubbles blown in the neutral hydrogen by the first stars.

We should also ask, `What if the EoR looks nothing like this?' There are possibilities which have received little consideration so far, including
Exactly what was happening in the dark sector during the dark ages ?
Is there a role for Dark stars? Or for Self Interacting Dark Matter?

The DECam EoR Deep Fields and their followup with upcoming higher resolution narrowfield facilities may shed light on
\begin{itemize}
\item Bulge properties of first-light galaxies 
\item High redshift AGN
\item SuperMassive Black Hole seeds
\item assembly of galaxy mass as a function of look-back time
\item Pair production SNe (massive stars) at M$_K$ = --23
\item Young globular clusters with 10$^6$ year free fall times and M/L approaching 10$^{-4}$
\end{itemize}

\section*{\it Acknowledgement}
The DECam deep field team also involves M.
Trenti, S. Wyithe, J. Cooke, C. Lidman, T. Abbott, A. Kunder, A. Koekemoer, E. Tescari, \& A. Katsianis.
Our DECam time to date was allocated by the Australian Time Allocation Committee.
There is a time exchange agreement between NOAO/CTIO and AAO which makes this possible.
AAO facilities are now reciprocally available to the NOAO user community and appear in the NOAO Newsletter in full detail.
Especially in demand and complementary to Tololo is the AAOmega MultiObject Spectrograph. 
CAASTRO supported my participation, the ARC's Centre of Excellence for All Sky Astrophysics.
We are indebted to all the people who made DECam the powerful and efficient machine that it is.
Happy Birthday, CTIO.

\section*{References}
\bibliography{editor}

\noindent Bahcall, J. \& Soneira, R. 1980, ApJS, 44, 73 (BS80)\\
Beckwith, S. et al 2006, AJ, 132, 1729\\
Bender, R. et al 2005, ApJ 631, 280 (B+05)\\
Cooke, J. et al 2012, Nature, 491, 228\\
Dilday, B. et al 2010, ApJ, 713, 1026\\
Gebhardt, K. et al 2001, AJ, 122, 2469 (G+01)\\
Ghez, A. et al 2005, ApJ 620, 744 (G+05)\\
Graham, A. 2001,     Rev Mex AA, 171, 97 (G01)\\ 
Green, J. et al 2011, astro-ph 1208.4012\\
Heger, A. \& Woosley, S. 2002, ApJ, 567, 532\\	
Heger, A. et al. 2012, ASP Conf Ser 458, 11\\	
Illingworth, G. et al 2013, astro-ph 1305.1931\\
Kormendy, J. \& Kennicutt, R. 2004, ARAA, 42, 36\\
Lawrence, J. et al 2009, PASA, 26, 379\\
Mould, J. 2011, PASA, 28, 266\\
Mould, J. 2013, PASA, 30, 27\\
Regan, M. \& Vogel, S., 1994, ApJ 434, 536 (RV94)\\
Spergel, D. et al 2013,  arXiv:1305.5425\\
Stetson, P. 1987, PASP, 99, 191\\
Taylor, A. 2013, IAU Symposium 291, 337\\
van der Marel, R. et al 1998, ApJ, 493, 613 (vdM+98)\\
Whalen, D. et al 2013, ApJ, 762, L6\\
Widrow, L., Perrett, K. \& Suyu, S. 2003, ApJ, 588, 311 (WPS03)\\  

\end{document}